\documentclass[11pt,titlepage,a4paper]{article}
\usepackage{bozhomac}	
\usepackage{bozhlogo}	
\usepackage{cite}	
\usepackage{tabularx}	
\usepackage{varioref}	

\title{\bfseries	\vspace*{-1.678902345in}
{\huge Equivalence principle in\\[1ex] classical electrodynamics}
}

\vspace{1.7ex}

\author{
Bozhidar Z.\ Iliev
\thanks{Laboratory of Mathematical Modeling in Physics,
Institute for Nuclear Research and \mbox{Nuclear} Energy,
Bulgarian Academy of Sciences,
Boul.\ Tzarigradsko chauss\'ee~72, 1784 Sofia, Bulgaria}
\thanks{E-mail address: bozho@inrne.bas.bg}
\thanks{URL: http://theo.inrne.bas.bg/$^\sim$bozho/}
}

\date{
\vspace{2.27ex}\ShortTitle{Equivalence principle in electrodynamics}\\[0.27ex]
\vspace{3.27ex}
\small
	\begin{tabular}{r@{$\colon\to~$}l}
 \vspace{0.27ex} Produced	& \fbox{\today}	\\[0.27ex]
	\end{tabular} \\[1.27ex]
\normalsize
	\begin{tabular}{r@{$\colon~$}l}
\vspace{0.27ex} http://www.arXiv.org e-Print archive No. & gr-qc/0303002
 								\\[0.27ex]
	\end{tabular} \\[-0.27ex]
 \vspace{4.27ex}{\Huge\BOZHO}	\\[4.27ex]
\vspace{0.27ex}\Subject{General relativity, Classical electrodynamics}
							\\[2.27ex]
	\begin{tabular}{r@{\hspace{0.512em}}|@{\hspace{0.512em}}l}
\vspace{0.27ex} 
&
\vspace{0.27ex}
	\end{tabular} \\[1.27ex]
\vspace{0.27ex}	\\[0.27ex]
}

\begin{document}		

\renewcommand{\thepage}{\roman{page}}

\renewcommand{\thefootnote}{\fnsymbol{footnote}} 
\maketitle				
\renewcommand{\thefootnote}{\arabic{footnote}}   

\tableofcontents		


\begin{abstract}

	The principle of equivalence in gravitational physics and its
mathematical base are reviewed. It is demonstrated how this principle can be
realized in classical electrodynamis. In general, it is valid at any given
single point or along a path without selfintersections unless the field
considered satisfies some conditions.

\end{abstract}

\renewcommand{\thepage}{\arabic{page}}


\section {Introduction}
\label{Introduction}

	The equivalence principle is a well known statement in gravitational
physics~\cite{MTW,Weinberg,JL-Anderson,Torretti,Synge,Moller,bp-PE-P?}.
The range of its validity was extended in~\cite{bp-NF-D+EP} to cover
classical gauge theories. The present paper concentrate to its realization in
classical electrodynamics.

	Section~\ref{Sect2} introduces the rigorous mathematical background
on which the equivalence principle is based. Sections~\ref{Subsect2.1} and
~\ref{Subsect2.2} contain a brief review of the concepts of linear connection
and linear transport along paths in vector bundles. Some links between these
objects are investigated in Sect.~\ref{Subsect2.3}. In Sect.~\ref{Subsect2.4}
are considered the so\ndash called normal frames for linear connections and
linear transports along paths in vector bundles. The importance of these
concepts for the physics comes from the fact that they turn to be the
mathematical equivalent to the physical notion of an inertial frame of
reference.

	In~\ref{Sect3} are reviewed some properties of the electromagnetic
potentials in classical electrodynamics. Special attention is paid to their
geometrical interpretation as (3\ndash index) coefficients of a linear
connection (parallel transport along paths) in one\ndash dimensional vector
bundle. The equivalence principle in gravity physics is considered in
Sect.~\ref{Sect4}. Section~\ref{Sect5} is devoted to the principle of
equivalence in classical electrodynamics.  As in a gravity theory (based on
linear connection(s)), it can be formulated as the assertion for coincidence
of inertial and normal frames for a given electromagnetic field. Similarly it
is identically valid at any single point or injective path in the spacetime
and on other subsets it does not hold generally if some additional conditions
are fulfilled.

	Section~\ref{Conclusion} ends the work with some concluding remarks.


\section
[Normal frames for linear transports along paths and linear connections]
{Normal frames for linear transports along paths and\\ linear connections}
\label{Sect2}

	The general theory of frames normal for a broad class of derivations,
in particular covariant derivatives (linear connections), and linear
transports along paths (which in particular can be parallel transports
generated by linear connections) is developed
in~\cite{bp-Frames-n+point,bp-Frames-path,bp-Frames-general,
bp-normalF-LTP,bp-NF-D+EP} and in the references therein. The material
in this section is abstracted from these works and concerns mainly linear
connections. Since the classical electromagnetic field is naturally described
via linear connections (see Sect.~\ref{Sect3}), this is done with the
intention for applying the general theory mentioned to the classical
electrodynamics (see Sect.~\ref{Sect5}).

\subsection{Linear connections in vector bundles}
\label{Subsect2.1}

	Different equivalent definitions of a (linear) connection in vector
bundles are known and in current usage~\cite{K&N,Poor, Greub&et_al.,
Bleecker}. The most suitable one for our purposes is given
in~\cite[p.~223]{Baez&Muniain} or~\cite[p.~281]{Eguchi&et_al.} (see
also~\cite[theorem~2.52]{Poor}).

	Suppose $(E,\pi,M)$, $E$ and $M$ being finite-dimensional  $C^\infty$
manifolds, be $C^\infty$ \field-vector bundle~\cite{Poor} with bundle space
$E$, base $M$, and projection $\pi\colon E\to M$. Here \field\ stands for the
field \field[R] of real numbers or \field[C] of complex ones. Let
$\Sec^k(E,\pi,M)$, $k=0,1,2,\dots$ be the set (in fact the module) of $C^k$
sections of $(E,\pi,M)$ and $\Fields{X}(M)$ the one of vector fields on $M$.

	\begin{Defn}	\label{Defn2.1}
	Let $V,W\in\Fields{X}(M)$, $\sigma,\tau\in\Sec^1(E,\pi,M)$, and
$f\colon M\to\field$ be a $C^\infty$ function. A mapping
 $\nabla\colon\Fields{X}(M)\times\Sec^1(E,\pi,M)\to\Sec^0(E,\pi,M)$,
 $\nabla\colon(V,\sigma)\mapsto\nabla_V\sigma$, is called a
\emph{(linear) connection in} $(E,\pi,M)$ if:
	\begin{subequations}		\label{2.1}
	\begin{align}	\label{2.1a}
\nabla_{V+W}\sigma & = \nabla_{V}\sigma + \nabla_{W}\sigma ,
\\			\label{2.1b}
\nabla_{fV}\sigma & = f \nabla_{V}\sigma ,
\\			\label{2.1c}
\nabla_{V}(\sigma+\tau) & = \nabla_{V}(\sigma) + \nabla_{V}(\tau) ,
\\			\label{2.1d}
\nabla_{V}(f\sigma) & = V(f)\cdot\sigma + f\cdot \nabla_{V}(\sigma) .
	\end{align}
	\end{subequations}
	\end{Defn}

	\begin{Rem}	\label{Rem2.1}
	Rigorously speaking, $\nabla$, as defined by
definition~\ref{Defn2.1}, is a covariant derivative operator in $(E,\pi,M)$
--- see~\cite[definition~2.51]{Poor} --- but, as a consequence
of~\cite[theorem~2.52]{Poor}, this cannot lead to some ambiguities.
	\end{Rem}

	\begin{Rem}	\label{Rem2.2}
	Since $V(a)=0$ for every $a\in\field$ (considered as a constant
function $M\to\{a\}$), the mapping
$\nabla\colon(V,\sigma)\mapsto\nabla_V\sigma$ is \field-linear with respect
to both its arguments.
	\end{Rem}

	Let $\{e_i: i=1,\dots,\dim\pi^{-1}(x)\}$, $x\in M$ and
$\{E_\mu:\mu=1,\dots,\dim M\}$ be frames over an open set  $U\subseteq M$
in, respectively, $(E,\pi,M)$ and the tangent bundle $(T(M),\pi_T,M)$ over
$M$, \ie for every $x\in U$, the set $\{e_i|_x\}$ forms a basis of the fibre
$\pi^{-1}(x)$ and $\{E_\mu|_x\}$ is a basis of the space
$T_x(M)=\pi_T^{-1}(x)$ tangent to $M$ at $x$. Let us write
$\sigma=\sigma^ie_i$ and $V=V^\mu E_\mu$, where here and henceforth the Latin
(resp.\ Greek) indices run from 1 to the dimension of $(E,\pi,M)$
(resp.\ $M$), the Einstein summation convention is assumed, and
$\sigma^i,V^\mu\colon U\to\field$ are some $C^1$ functions. Then, from
definition~\ref{Defn2.1}, one gets
	\begin{equation}	\label{2.2}
\nabla_V\sigma =
  V^\mu\bigl(
	E_\mu(\sigma^i) + \Sprindex[\Gamma]{j\mu}{i} \sigma^j
  \bigr) e_i
	\end{equation}
where $\Sprindex[\Gamma]{j\mu}{i}\colon U\to\field$, called
\emph{coefficients} of $\nabla$, are given by
	\begin{equation}	\label{2.3}
\nabla_{E_\mu}e_j =: \Sprindex[\Gamma]{j\mu}{i} e_i .
	\end{equation}

	Evidently, by virtue of~\eref{2.2}, the knowledge of
$\{\Sprindex[\Gamma]{j\mu}{i}\}$ in a pair of frames $(\{e_i\},\{E_\mu\})$
over $U$ is equivalent to the one of $\nabla$ as any transformation
\(
(\{e_i\},\{E_\mu\})
  \mapsto
	(\{e'_i=A_i^je_j\} , \{E'_\mu=B_\mu^\nu E_\nu\})
\)
with non\ndash degenerate matrix\ndash valued functions $A=[A_i^j]$ and
$B=[B_\mu^\nu]$ on  $U$ implies
\(
\Sprindex[\Gamma]{j\mu}{i}
  \mapsto
	\Sprindex[\Gamma]{j\mu}{\prime\mspace{0.7mu} i}
\)
with
	\begin{equation}	\label{2.4}
\Sprindex[\Gamma]{j\mu}{\prime\mspace{0.7mu} i}
  = \sum_{\nu=1}^{\dim M} \sum_{k,l=1}^{\dim\pi^{-1}(x)}
	B_{\mu}^{\nu} \bigl(A^{-1}\bigr)_{k}^{i} A_{j}^{l}
		\Sprindex[\Gamma]{l\nu}{k}
	+ \sum_{\nu=1}^{\dim M} \sum_{k=1}^{\dim\pi^{-1}(x)}
		B_{\mu}^{\nu}\bigl(A^{-1}\bigr)_{k}^{i}
		E_\nu (A_{j}^{k})  .
	\end{equation}
which in a matrix form reads
	\begin{equation}	\label{2.5}
{\Gamma}_\mu^{\prime}
  = B_{\mu}^{\nu} A^{-1}{\Gamma}_\nu A  + A^{-1} E'_\mu(A)
  = B_{\mu}^{\nu} A^{-1}\bigl( {\Gamma}_\nu A  + E_\nu(A) \bigr)
	\end{equation}
where
$\Gamma_\mu:=[\Sprindex[\Gamma]{j\mu}{i}]_{i,j=1}^{\dim\pi^{-1}(x)}$,
$x\in M$, and
\(
\Gamma'_\mu
	:=[\Sprindex[\Gamma]{j\mu}{\prime\,i}]_{i,j=1}^{\dim\pi^{-1}(x)} .
\)

\subsection{Linear transports along paths in vector bundles}
\label{Subsect2.2}

	To begin with, we recall some definitions and results from the
paper~\cite{bp-normalF-LTP}.%
\footnote{%
In~\cite{bp-normalF-LTP} is assumed $\field=\field[C]$ but this choice is
insignificant.%
}
Below we denote by $\PLift^k(E,\pi,M)$ the set of liftings of $C^k$ paths
from $M$ to $E$ such that the lifted paths are of class $C^k$,
$k=0,1,\ldots$. Let $\gamma\colon J\to M$, $J$ being real interval, be a path
in M.

	\begin{Defn}	\label{Defn3.1}
	A \emph{linear transport along paths in vector bundle} $(E,\pi,M)$ is
a mapping $L$ assigning to every path $\gamma$ a mapping $L^\gamma$,
\emph{transport along} $\gamma$, such that
$L^\gamma\colon (s,t)\mapsto L^\gamma_{s\to t}$ where the mapping
	\begin{equation}	\label{3.1}
L^\gamma_{s\to t} \colon  \pi^{-1}(\gamma(s)) \to \pi^{-1}(\gamma(t))
	\qquad s,t\in J,
	\end{equation}
called \emph{transport along $\gamma$ from $s$ to} $t$, has the properties:
	\begin{alignat}{2}	\label{3.2}
L^\gamma_{s\to t}\circ L^\gamma_{r\to s} &=
			L^\gamma_{r\to t},&\qquad  r,s,t&\in J, \\
L^\gamma_{s\to s} &= \id_{\pi^{-1}(\gamma(s))}, & s&\in J,	\label{3.3}
\\
L^\gamma_{s\to t}(\lambda u + \mu v) 				\label{3.4}
  &= \lambda L^\gamma_{s\to t}u + \mu L^\gamma_{s\to t}v,
	& \lambda,\mu &\in \field,\quad u,v\in{\pi^{-1}(\gamma(s))},
	\end{alignat}
where  $\circ$ denotes composition of maps and
$\id_X$ is the identity map of a set $X$.
	\end{Defn}

	\begin{Defn}	\label{Defn3.2}
	A \emph{derivation along paths in} $(E,\pi,M)$ or a
\emph{derivation of liftings of paths in} $(E,\pi,M)$ is a mapping
	\begin{subequations}	\label{3.5}
	\begin{equation}	\label{3.5a}
	D\colon\PLift^1(E,\pi,M) \to \PLift^0(E,\pi,M)
	\end{equation}
	\end{subequations}
which is \field-linear,
	\begin{subequations}	\label{3.6}
	\begin{equation}	\label{3.6a}
D(a\lambda+b\mu) = aD(\lambda) + bD(\mu)
	\end{equation}
	\end{subequations}
for $a,b\in\field$ and $\lambda,\mu\in\PLift^1(E,\pi,M)$, and the mapping
	\begin{equation}
	\tag{\ref{3.5}b}	\label{3.5b}
D_{s}^{\gamma}\colon \PLift^1(E,\pi,M) \to \pi^{-1}(\gamma(s)),
	\end{equation}
defined via
\(
D_{s}^{\gamma}(\lambda)
 := \bigl( (D(\lambda))(\gamma) \bigr) (s)
  = (D\lambda)_\gamma(s)
\)
and called \emph{derivation along $\gamma\colon J\to M$ at} $s\in J$,
satisfies the `Leibnitz rule':
	\begin{equation}
	\tag{\ref{3.6}b}	\label{3.6b}
D_s^\gamma(f\lambda)
 = \frac{\od f_\gamma(s)}{\od s} \lambda_\gamma(s)
	+ f_\gamma(s) D_s^\gamma(\lambda)
	\end{equation}
for every
\[
f\in
\PF^1(M) :=
	\{
	\varphi|\varphi\colon\gamma\mapsto\varphi_\gamma, \
\gamma\colon J\to M,\ \varphi_\gamma\colon J\to \field
	\text{ being of class $C^1$}
	\} .
\]
The mapping
	\begin{equation}
	\tag{\ref{3.5}c}	\label{3.5c}
D^{\gamma}\colon \PLift^1(E,\pi,M) \to
	\Path\bigl(\pi^{-1}(\gamma(J))\bigr)
	:=\{ \text{paths in } \pi^{-1}(\gamma(J)) \} ,
	\end{equation}
defined by $D^\gamma(\lambda):=(D(\lambda))|_\gamma=(D\lambda)_\gamma$, is
called \emph{derivation along} $\gamma$.
	\end{Defn}

	If $\gamma\colon J\to M$ is a path in $M$ and $\{e_i(s;\gamma)\}$ is
a basis in $\pi^{-1}(\gamma(s))$,%
\footnote{%
If there are $s_1,s_2\in J$ such that $\gamma(s_1)=\gamma(s_2):=y$, the
vectors $e_i(s_1;\gamma)$ and $e_i(s_2;\gamma)$ need not to coincide. So, if
this is the case, the bases $\{e_i(s_1;\gamma)\}$ and $\{e_i(s_2;\gamma)\}$
in $\pi^{-1}(y)$ may turn to be different.%
}
in the frame $\{e_i\}$ over $\gamma(J)$ the
\emph{components (matrix elements)} $\Sprindex[L]{j}{i}\colon U\to\field$ of
a linear transport $L$ along paths and the ones of a derivation $D$ along
paths in vector bundle $(E,\pi,M)$ are defined through, respectively,
	\begin{gather}	\label{3.7}
L_{s\to t}^{\gamma} \bigl(e_i(s;\gamma)\bigr)
		=:\Sprindex[L]{i}{j}(t,s;\gamma) e_j(t;\gamma)
		\qquad s,t\in J,
\\			\label{3.8}
D_s^\gamma \hat{e}_j
=: \Sprindex[\Gamma]{j}{i}(s;\gamma) e_i(s;\gamma)
\qquad s\in J
	\end{gather}
where $\hat{e}_i\colon\gamma\mapsto e_i(\cdot;\gamma)$ is a lifting of
$\gamma$ generated by $e_i$.

	It is a simple exercise to verify that the components of $L$ and $D$
uniquely define (locally) their action on $u=u^ie_i(s;\gamma)$ and
$\lambda\in\PLift^1(E,\pi,M)$,
$\lambda\colon\gamma\mapsto\lambda_\gamma=\lambda_\gamma^i\hat{e}_i$,
according to
	\begin{gather}	\label{3.9}
L_{s\to t}^{\gamma} u
	=:\Sprindex[L]{j}{i}(t,s;\gamma) u^j e_i(t;\gamma)
\\			\label{3.10}
D_s^\gamma \lambda
	=: \Bigl(
	   \frac{\od\lambda_\gamma^i(s)}{\od s}
	   +
	   \Sprindex[\Gamma]{j}{i}(s;\gamma) \lambda_\gamma^j(s)
	   \Bigr) e_i(s;\gamma)
	\end{gather}
and that a change
$\{e_i(s;\gamma)\} \mapsto \{e'_i(s;\gamma)=A_i^j(s;\gamma)e_j(s;\gamma)\}$,
with a non\ndash degenerate matrix\ndash valued function
 $A(s;\gamma) := [A_i^j(s;\gamma)]$, implies the transformations
	\begin{gather}	\label{3.11}
\Mat{L}(t,s;\gamma) :=\bigl[ \Sprindex[L]{i}{j}(t,s;\gamma) \bigr]
\mapsto
\Mat{L}^\prime(t,s;\gamma)
  = A^{-1}(t;\gamma) \Mat{L}(t,s;\gamma) A(s;\gamma)
\displaybreak[1]\\	\label{3.12}
\Mat{\Gamma}(s;\gamma) := \bigl[ \Sprindex[\Gamma]{j}{i}(s;\gamma) \bigr]
\mapsto
\Mat{\Gamma}^\prime(s;\gamma)
  = A^{-1}(s;\gamma) \Mat{\Gamma}(s;\gamma) A(s;\gamma)
    + A^{-1}(s;\gamma)\frac{\od A(s;\gamma)}{\od s}.
	\end{gather}

	A crucial role further will be played by the \emph{coefficients}
$\Sprindex[\Gamma]{j}{i}(s;\gamma)$ in a frame $\{e_i\}$ of  linear
transport $L$,
	\begin{equation}	\label{3.13}
\Sprindex[\Gamma]{j}{i}(s;\gamma)
   := \frac{\pd\Sprindex[L]{j}{i}(s,t;\gamma)}{\pd t}\bigg|_{t=s}
   = - \frac{\pd\Sprindex[L]{j}{i}(s,t;\gamma)}{\pd s}\bigg|_{t=s} .
	\end{equation}
The usage of the same notation for the \emph{coefficients} of a
transport $L$ and \emph{components} of derivation $D$ along paths is not
accidental and finds its reason in the following fundamental
result~\cite[sec.~2]{bp-normalF-LTP}. Call a transport $L$ differentiable of
class $C^k$, $k=0,1,\dots$ if its matrix $\Mat{L}(t,s;\gamma)$ has $C^k$
dependence on $t$ (and hence on $s$ --- see~\cite[sec.~2]{bp-normalF-LTP}).
\emph{Every $C^1$ linear transport $L$ along paths generates a derivation $D$
along paths via
	\begin{equation}	\label{3.14}
D_{s}^{\gamma}(\lambda)
   := \lim_{\varepsilon\to0}\Bigl\{\frac{1}{\varepsilon}\bigl[
       L_{s+\varepsilon\to s}^{\gamma}\lambda_\gamma(s+\varepsilon)
       - \lambda_\gamma(s)\bigr]\Bigr\}
	\end{equation}
for every lifting $\lambda\in\PLift^1(E,\pi,M)$
with $\lambda\colon\gamma\mapsto\lambda_\gamma$
and conversely, for any derivation $D$ along paths there exists a unique
linear transport along paths generating it via~\eref{3.14}.}
Besides,
\emph{if $L$ and $D$ are connected via~\eref{3.14}, the coefficients of
$L$ coincide with the components of $D$}.
In short, there is a bijective
correspondence between linear transports and derivations along paths given
locally through the equality of their coefficients and components
respectively.

	More details and results on the above items can be found
in~\cite{bp-normalF-LTP}.

\subsection
[Links between linear connections and linear transports]
{Links between linear connections and linear transports}
\label{Subsect2.3}

	Suppose $\gamma\colon J\to M$ is a  $C^1$ path and $\dot\gamma(s)$,
$s\in J$, is the vector tangent to $\gamma$ at $\gamma(s)$ (more precisely, at
$s)$. Let $\nabla$ and $D$ be, respectively, a linear connection and
derivation along paths in vector bundle $(E,\pi,M)$, and in a pair of frames
$(\{e_i\},\{E_\mu\})$ over some open set in $M$ the coefficients of $\nabla$
and the components of $D$ be $\Sprindex[\Gamma]{j\mu}{i}$ and
$\Sprindex[\Gamma]{j}{i}$ respectively, \ie
 $\nabla_{E_\mu} e_i = \Sprindex[\Gamma]{i\mu}{j} e_j$
and
 $D_s^\gamma \hat{e}_i = \Sprindex[\Gamma]{i}{j} e_j(\gamma(s))$
with
$\hat{e}_i\colon\gamma\mapsto\hat{e}_i|_\gamma\colon s\mapsto e_i(\gamma(s))$
being lifting of paths generated by $e_i$. If
$\sigma=\sigma^ie_i\in\Sec^1(E,\pi,M)$ and $\hat\sigma\in\PLift(E,\pi,M)$ is
given via
$\hat\sigma\colon\gamma\mapsto\hat\sigma_\gamma:=\sigma\circ\gamma$,
then~\eref{3.10} implies
	\begin{equation*}
D^\gamma_s \hat\sigma
= \Big(
	\frac{\od\sigma^i(\gamma(s))}{\od s}
	+
	\Sprindex[\Gamma]{j}{i}(s;\gamma) \sigma^j(\gamma(s))
  \Bigr) e_i(\gamma(s))
	\end{equation*}
while, if $\gamma(s)$ is not a self-intersection point for
$\gamma$, equation~\eref{2.2} leads to
	\begin{equation*}
(\nabla_{\dot\gamma}\sigma)|_{\gamma(s)}
= \Big(
	\frac{\od\sigma^i(\gamma(s))}{\od s}
	+
	\Sprindex[\Gamma]{j\mu}{i}(\gamma(s))
  				  \sigma^j(\gamma(s))\dot{\gamma}^\mu(s)
  \Bigr) e_i(\gamma(s)) .
	\end{equation*}
Obviously, we have
	\begin{equation}	\label{4.1}
D^\gamma_s \hat\sigma = (\nabla_{\dot\gamma}\sigma)|_{\gamma(s)}
	\end{equation}
for every $\sigma$ iff
	\begin{equation}	\label{4.2}
\Sprindex[\Gamma]{j}{i}(s;\gamma)
=
\Sprindex[\Gamma]{j\mu}{i}(\gamma(s)) \dot\gamma^\mu(s)
	\end{equation}
which, in matrix form reads
	\begin{equation}	\label{4.3}
\Mat{\Gamma}(s;\gamma) = \Gamma_\mu(\gamma(s))\dot\gamma^\mu(s) .
	\end{equation}
A simple algebraic calculation shows that this equality is invariant under
changes of the frames $\{e_i\}$ in $(E,\pi,M)$ and $\{E_\mu\}$ in
$(T(M),\pi_T,M)$. Besides, if~\eref{4.2} holds, then $\Mat{\Gamma}$
transforms according to~\eref{3.12} iff $\Gamma_\mu$ transforms according
to~\eref{2.5}.

	The above considerations are a hint that the linear connections
should, and in fact can, be described in terms of derivations or,
equivalently, linear transports along paths; the second description being
more relevant if one is interested in the parallel transports generated by
connections.

	\begin{Thm}	\label{Thm4.1}
	If $\nabla$ is a linear connection, then there exists a derivation $D$
along paths such that~\eref{4.1} holds for every $C^1$ path
$\gamma\colon J\to M$ and every $s\in J$ for which $\gamma(s)$ is not
self\ndash intersection point for $\gamma$.%
\footnote{%
In particular, $\gamma$ can be injective and $s$ arbitrary. If we restrict
the considerations to injective paths, the derivation $D$ is unique. The
essential point here is that at the self\ndash intersection points of
$\gamma$, if any, the mapping
$\dot\gamma\colon\gamma(s)\mapsto\dot\gamma(s)$
is generally multiple\ndash valued and, consequently, it is not a vector
field (along $\gamma$); as a result
$(\nabla_{\dot\gamma}\sigma)|_{\gamma(s)}$ at them becomes also
multiple\ndash valued.%
}
The matrix of the components of $D$ is given by~\eref{4.3} for every $C^1$
path $\gamma\colon J\to M$ and  $s\in J$ such that $\gamma(s)$ is not a
self\ndash intersection point for $\gamma$. Conversely, given a derivation
$D$ along path whose matrix along any $C^1$ path $\gamma\colon J\to M$ has
the form~\eref{4.3} for some matrix\ndash valued functions $\Gamma_\mu$,
there is a unique linear connection $\nabla$ whose matrices of coefficients
are exactly $\Gamma_\mu$ and for which, consequently,~\eref{4.1} is valid at
the not self\ndash intersection points of $\gamma$.
	\end{Thm}

	\begin{Proof}
	NECESSITY.
	If $\Gamma_\mu$ are the matrices of the coefficients of $\nabla$ in
some pair of frames $(\{e_i\},\{E_\mu\})$, define the matrix $\Mat{\Gamma}$
of the components of $D$ via~\eref{4.3} for any $\gamma\colon J\to M$.
	SUFFICIENCY. Given $D$ for which the decomposition~\eref{4.3} holds
in $(\{e_i\},\{E_\mu\})$ for any $\gamma$. It is trivial to verity that
$\Gamma_\mu$ transform according to~\eref{2.5} and, consequently, they are
the matrices of the coefficients of a linear connection $\nabla$ for which,
evidently,~\eref{4.1} holds.
	\end{Proof}

	A trivial consequence of the above theorem is the following important
result.

	\begin{Cor}	\label{Cor4.1}
	There is a bijective correspondence between the set of linear
connections in a vector bundle and the one of derivations along paths in it
whose components' matrices admit (locally) the decomposition~\eref{4.3}.
Locally, along a $C^1$ path $\gamma$ and pair of frames $(\{e_i\},\{E_\mu\})$
along it, it is given by~\eref{4.3} in which $\Mat{\Gamma}$ and $\Gamma_\mu$
are the matrices of the components of a derivation along paths and of the
coefficients of a linear connection, respectively.
	\end{Cor}

	Let us now look on the preceding material from the view-point of
linear transports along paths and parallel transports generated by linear
connections.

	Recall (see, e.g.,~\cite[chapter~2]{Poor}), a section
$\sigma\in\Sec^1(E,\pi,M)$ is \emph{parallel along}  $C^1$ path
$\gamma\colon J\to M$ with respect to a linear connection $\nabla$ if
$(\nabla_{\dot\gamma}\sigma)|_{\gamma(s)}=0$, $s\in J$.%
\footnote{%
If $\gamma$ is not injective, here and henceforth
$(\nabla_{\dot\gamma}\sigma)|_{\gamma(s)}$
should be replaced by $D_s^\gamma\hat\sigma$,
$\hat\sigma\colon\gamma\mapsto\sigma\circ\gamma$, where $D$ is the derivation
along paths corresponding to $\nabla$ via corollary~\ref{Cor4.1}.%
}
The \emph{parallel transport along} a $C^1$ path $\alpha\colon[a,b]\to M$,
$a,b\in\field[R]$, $a\le b$, generated by $\nabla$ is a mapping
\[
P^\alpha\colon \pi^{-1}(\alpha(a))\to \pi^{-1}(\alpha(b))
\]
such that $P^\alpha(u_0):=u(b)$ for every element $u_0\in\pi^{-1}(\alpha(a))$
where $u\in\Sec^1(E,\pi,M)|_{\alpha([a,b])}$ is the unique solution of the
initial\ndash value problem
	\begin{equation}	\label{4.4}
\nabla_{\dot\alpha} u = 0,
\qquad
u(a)=u_0 .
	\end{equation}
The \emph{parallel transport} $P$ generated by (assigned to, corresponding
to) a linear connection $\nabla$ is a mapping assigning to any
$\alpha\colon[a,b]\to M$ the parallel transport $P^\alpha$ along $\alpha$
generated by $\nabla$.

	Let $D$ be the derivation along paths corresponding to $\nabla$
according to corollary~\ref{Cor4.1}. Then~\eref{4.1} holds for
$\gamma=\alpha$, so~\eref{4.4} is tantamount to
	\begin{equation}	\label{4.5}
D_s^\alpha \hat u = 0
\qquad
u(a) = u_0
	\end{equation}
where $\hat u\colon\alpha\mapsto \bar u\circ\alpha$ with $\bar
u\in\Sec^1(E,\pi,M)$ such that $\bar u|_{\alpha([a,b])}=u$. From here and the
results of~\cite[sec.~2]{bp-normalF-LTP} immediately follows that the lifting
$\hat u$ is generated by the unique linear transport $\Psf$ along paths
corresponding to  $D$,
	\begin{equation}	\label{4.6}
\hat u\colon\alpha\mapsto
\hat{u}_\alpha := \bar{\Psf}_{a,u_0}^{\alpha},
\quad
\bar{\Psf}_{a,u_0}^{\alpha}\colon s\mapsto
\bar{\Psf}_{a,u_0}^{\alpha}(s):= \Psf_{a\to s}^{\alpha} u_0,
\qquad
s\in [a,b] .
	\end{equation}
Therefore
\(
P^\alpha(u_o)
:= u(b)
= \bar u(\alpha(b))
= \hat{u}_\alpha(b)
= \Psf_{a\to b}^{\alpha} u_0 .
\)
Since this is valid for all $u_0\in\pi^{-1}(\alpha(a))$, we have
	\begin{equation}	\label{4.7}
P^\alpha = \Psf_{a\to b}^{\alpha} .
	\end{equation}

	\begin{Thm}	\label{Thm4.2}
	The parallel transport $P$ generated by a linear connection $\nabla$
in a vector bundle coincides, in a sense of~\eref{4.7}, with the unique linear
transport $\Psf$ along paths in this bundle corresponding to the derivation
$D$ along paths defined, via corollary~\ref{Cor4.1}, by the connection.
Conversely, if $\Psf$ is a linear transport along paths whose coefficients'
matrix admits the representation~\eref{4.3}, then for every $s,t\in[a,b]$
	\begin{equation}	\label{4.8}
\Psf_{s\to t}^{\alpha} =
\begin{cases}
P^{\alpha|[s,t]}			&\text{for $s\le t$}	\\
\bigl(P^{\alpha|[t,s]}\bigr)^{-1}	&\text{for $s\ge t$}
\end{cases} ,
	\end{equation}
where $P$ is the parallel transport along paths generated by the unique
linear connection $\nabla$ corresponding to the derivation $D$ along paths
defined by $\Psf$.
	\end{Thm}

	\begin{Proof}
	The first part of the assertion was proved above while
deriving~\eref{4.7}. The second part is simply the inversion of all logical
links in the first one, in particular~\eref{4.8} is the solution
of~\eref{4.7} with respect to $\Psf$.
	\end{Proof}

	The transport $\Psf$ along paths corresponding according to
theorem~\ref{Thm4.2} to a parallel transport $P$ or a linear connection
$\nabla$ will be called \emph{parallel transport along paths}.

	\begin{Cor}	\label{Cor4.2}
	The local coefficients' matrix $\Gamma$ of a parallel transport along
paths and the coefficients' matrices $\Gamma_\mu$ of the generating it (or
generated by it) linear connection are connected via~\eref{4.3} for every
$C^1$ path $\gamma\colon J\to M$.
	\end{Cor}

	\begin{Proof}
	See theorem~\ref{Thm4.2}.
	\end{Proof}

	If the coefficients of a linear transport along paths admit a
representation~\eref{4.3} for any $\gamma\colon J\to U\subseteq M$, we shall
call $\Sprindex[\Gamma]{j\mu}{i}\colon U\to\field[K]$ its
3\ndash\emph{index coefficients}, $\Gamma=[\Sprindex[\Gamma]{j\mu}{i}]$ its
\emph{coefficient matrices}, and say that it admits 3\ndash index
coefficients.

	As there is a bijective correspondence between linear transports and
derivation along paths (locally given via the coincidence of their respective
coefficients and components --- see~\cite{bp-normalF-LTP}), from
corollary~\ref{Cor4.2} we get the following result.

       \begin{Cor}	\label{Cor4.2new}
A linear transport along paths admits 3-index coefficients on an \emph{open}
set $U\subseteq M$ if and only if it is a parallel transport along paths.
	\end{Cor}

	Notice, if $U\subset M$ is \emph{not} an open set in $M$, \eg if it
is a submanifold of dimension less than  $\dim M$, than
corollary~\ref{Cor4.2new} is generally not valid. The reason for that
conclusion is in that, if a transport admits 3\ndash index coefficients
$\Sprindex[\Gamma]{j\mu}{i}$ on $U$, then
$\Sprindex[\Gamma]{j\mu}{i}+\Sprindex[G]{j\mu}{i}$ are also its 3\ndash index
coefficients for any $\Sprindex[G]{j\mu}{i}\colon U\to\field[K]$ such that
$\Sprindex[G]{j\mu}{i}(\gamma(s))\dot\gamma^\mu(s)=0$ for any $C^1$ path

$\gamma\colon J\to U$. Consequently, we can assert that
$\Sprindex[G]{j\mu}{i}=0$ if $\dot\gamma(s)$ is an arbitrary vector in
$T_{\gamma(s)}(M)$, which is the case when $U$ is an open set in $M$; if $U$
is a submanifold and $\dim U<\dim M$, then $\Sprindex[G]{j\mu}{i}V^\mu=0$
with $V_x\in T_x(U)$ does not imply $\Sprindex[G]{j\mu}{i}=0$ for all
$\mu=1,\dots,\dim M$. So, generally the 3\ndash index coefficients of a
linear transport along paths, if any, are not defined uniquely, contrary to
the case of parallel transports along paths.

\subsection{Normal and strong normal frames}
\label{Subsect2.4}

	Freely speaking, a normal frame for a derivation (\eg linear
connection) or transport along paths (\eg parallel one) is a (local) frame in
the bundle space in which it looks (locally) as if we are dealing with an
ordinary derivation or parallel transport, respectively, in Euclidean space,
\ie in which it looks (locally) Euclidean. That intuitive understanding is
formalized in the following definitions in which we restrict ourselves to
linear connections due to further considerations in the present work.

	\begin{Defn}	\label{Defn5.1}
	Given a linear connection $\nabla$ in a vector bundle $(E,\pi,M)$ and
a subset $U\subseteq M$. A \emph{frame} $\{e_i\}$ in $E$ defined over an open
subset $V$ of $M$ containing $U$ or equal to it, $V\supseteq U$, is called
\emph{normal for $\nabla$ over} $U$ if in it and some (and hence any) frame
$\{E_\mu\}$ in $T(M)$ over $V$ the coefficients of $\nabla$ vanish everywhere
on $U$. Respectively, $\{e_i\}$ is \emph{normal for $\nabla$ along a mapping}
$g\colon Q\to M$, $Q\not=\varnothing$, if $\{e_i\}$ is normal for $\nabla$
over $g(Q)$.
	\end{Defn}

	\begin{Defn}	\label{Defn5.2}
	Given a linear transport $L$ (resp.\ derivation $D$) along paths in a
vector bundle $(E,\pi,M)$ and a subset $U\subseteq M$. A frame $\{e_i\}$ in
$E$ defined on an open set $V$ containing $U$, $V\supseteq U$, is called
\emph{normal} for $L$ (resp.\ $D$) on $U$ if in it vanish the coefficients of
$L$ (resp.\ components of $D$) along every path $\gamma\colon J\to U$. A frame
is called normal (along a path $\gamma\colon J\to M$) for $L$ or $D$ if it is
normal for it on $U=M$ (resp.\ $U=\gamma(J)$).
	\end{Defn}

	A linear connection or transport/derivation along paths is called
\emph{Euclidean} on $U\subseteq M$ if it admits a frame normal for it on $U$.

	A necessary condition for a linear transport along paths to be
Euclidean is provided by the following
result~\cite[proposition~5.1]{bp-normalF-LTP}.

	\begin{Prop}	\label{4-Prop5.1}
	For every Euclidean on $U\subseteq M$
(resp.\ along a $C^1$ path $\gamma\colon J\to M$)
linear transport $L$ along paths in $(E,\pi,M)$, $E$ and $M$ being $C^1$
manifolds, the matrix $\Mat{\Gamma}$ of its coefficients has the
representation
	\begin{equation}	\label{4-5.1}
\Mat{\Gamma}(s;\gamma)
   = \sum_{\mu=1}^{\dim M} \Gamma_\mu(\gamma(s)) \dot\gamma^\mu(s)
   \equiv \Gamma_\mu(\gamma(s)) \dot\gamma^\mu(s)
	\end{equation}
in any frame $\{e_i\}$ along every (resp.\ the given) $C^1$ path
$\gamma\colon J\to U$, where
\(
{\Gamma}_\mu =
  \bigl[\Sprindex[\Gamma]{j\mu}{i}\bigr]_{i,j=1}^{\dim\pi^{-1}(x)}
\)
are some matrix-valued functions, defined on an open set~$V$ containing $U$
(resp.\ $\gamma(J)`$) or equal to it, and $\dot\gamma^\mu$ are the components
of $\dot\gamma$ in some frame $\{E_\mu\}$ along $\gamma$ in the bundle space
$T(M)$ tangent to $M$, $\dot\gamma=\dot\gamma^\mu E_\mu$.
	\end{Prop}

	Combining this result with corollary~\ref{Cor4.2}, we see that the
parallel transports along paths \emph{may} admit normal frames. However, the
existence of such frames depends on the subset $U$ on which they are normal.
In particular, normal frames always exist if $U$ is a single point and for
$U=\gamma(J)$ for some path $\gamma\colon J\to M$; besides, the normal
frames are generally anholonomic. For instance, a linear transport $L$ is
Euclidean on $U\subseteq M$ iff it is path\ndash independent in $U$, \ie iff
$L_{s\to t}^{\gamma}$ depends only on the points $\gamma(s)$ and $\gamma(t)$
but not on the particular path in $U$ connecting them, or iff its matrix  in
a frame $\{e_i\}$ in $E$ is of the form
$\bs{L}(t,s)=\bs{F}_0^{-1}(\gamma(t))\bs{F}_0(\gamma(s))$
for $\gamma\colon J\to U$ and some non\ndash degenerate matrix\ndash valued
function $\bs{F}_0$ on $U$, or iff its coefficient's matrix in the same frame
is
\(
\Gamma(s;\gamma)
=\bs{F}_0^{-1}(\gamma(s)) \frac{\od\bs{F}_0(\gamma(s))}{\od s}.
\)
For details concerning existence,
uniqueness and holonomicity of frames normal for linear transports, the reader
is referred to~\cite{bp-normalF-LTP}.

	Since in this paper we shall be interested mainly in linear
connections, the next considerations will be restricted to frames normal for
linear connections and parallel transports along paths (generated by them).

	Let $\nabla$ and $\mathsf{P}$ be, respectively, a linear connection
on $M$ and the parallel transport along paths in $(E,\pi,M)$ generated by
$\nabla$ (see~\eref{4.2}). Suppose $\nabla$ and $\mathsf{P}$ admit frames
normal on a set $U\subseteq M$. Here a natural question arises:  what are the
links between both types of normal frames, the ones normal for $\nabla$ on
$U$ and the ones for $\mathsf{P}$ on $U$?

	Recall, if $\Sprindex[\Gamma]{jk}{i}$ are the components of $\nabla$
in a frame $\{E_i\}$, the frame $\{E_i\}$ is normal on $U\subseteq M$ for
$\nabla$ or $\mathsf{P}$ iff respectively
	\begin{gather}	\label{4-12.1}
\Sprindex[\Gamma]{jk}{i}(p) = 0
		\\	\label{4-12.2}
\Sprindex[\Gamma]{j}{i} (s;\gamma)
	= \Sprindex[\Gamma]{jk}{i}(\gamma(s)) \dot\gamma^k(s)
= 0
	\end{gather}
for every $p\in U$, $\gamma\colon J\to U$, and $s\in J$. From these
equalities two simple but quite important conclusions can be made: (i) The
frames normal for $\nabla$ are normal for $\mathsf{P}$, the opposite being
generally not valid, and (ii) in a frame normal for $\nabla$
vanish the 2\ndash index as well as the 3\ndash index coefficients of
$\mathsf{P}$.

	\begin{Defn}	\label{4-Defn12.1}
	Let $\mathsf{P}$ be a parallel transport in $(E,\pi,M)$ and
$U\subseteq M$. A \emph{frame} $\{E_i\}$, defined on an open set containing
$U$, is called \emph{strong normal on $U$ for} $\mathsf{P}$ if
the 3\ndash index coefficients of $\mathsf{P}$ in $\{E_i\}$ vanish on $U$.
Respectively, $\{E_i\}$ is \emph{strong normal along} $g\colon Q\to M$ if it
is strong normal on $g(Q)$.
	\end{Defn}

	Obviously, the set of frames strong normal on $U$ for a parallel
transport $\mathsf{P}$ coincides with the set of frames normal for the linear
connection $\nabla$ generating $\mathsf{P}$.

	The above considerations can be generalized  directly to linear
transports for which 3\ndash index coefficients exist and are fixed
(see~\cite[sec.~7]{bp-normalF-LTP}).

	As a sufficient criterion for existence of (strong) normal frames for
(a parallel transport generated by) linear connection, we shall present the
following result~\cite[theorem~10.1]{bp-NF-D+EP}

	\begin{Thm}	\label{Thm5.1}
	If $\gamma_n\colon J^n\to M$, $J^n$ being neighborhood in
$\mathbb{R}^n$, $n\in\mathbb{N}$, $n\le \dim M$, is a $C^1$ injective mapping,
then a necessary and sufficient condition for the existence of frame(s)
normal over $\gamma_n(J^n)$ for some linear connection in a vector bundle
$(E,\pi,M)$ is, in some neighborhood (in $\mathbb{R}^n$) of every $s\in J^n$,
their (3\ndash index) coefficients to satisfy the equations
	\begin{equation}	\label{5.2}
\bigl(
  R_{\mu\nu} (-\Gamma_1\circ\gamma_n,\ldots,-\Gamma_{\dim M}\circ\gamma_n)
\bigr)
	(s)
  = 0,
\qquad \mu,\nu=1,\ldots,n
	\end{equation}
where $R_{\mu\nu}$
(in a coordinate frame
$\bigl\{E_\mu=\frac{\pd}{\pd x^\mu}\bigr\}$ in a neighborhood of
$\gamma_n(s)\in M$) are given via
	\begin{multline}	\label{5.3}
R_{\mu\nu} (-\Gamma_1\circ\gamma_n,\ldots,-\Gamma_{\dim M}\circ\gamma_n)
\\
  := - \frac{\pd \Gamma_\mu(\gamma_n(s)) }{\pd s^\nu}
     + \frac{\pd \Gamma_\nu(\gamma_n(s)) }{\pd s^\mu}
     +  {\Gamma}_\mu(\gamma_n(s)) {\Gamma}_\nu(\gamma_n(s))
     -  {\Gamma}_\nu(\gamma_n(s)) {\Gamma}_\mu(\gamma_n(s)) .
	\end{multline}
for $\mu,\nu=1,\dots,n$. Here $\{s^1,\dots,s^n\}$ are Cartesian coordinates
in $\mathbb{R}^n$ and the local coordinates $\{x^\mu\}$ on $M$ are such that
$x(\gamma_n(s)) = (s,\bs{t}_0)$ for some fixed
$\bs{t}_0\in\field[R]^{\dim M - n }$ and in a neighborhood of
$\gamma_n(s)$ in $M$ the coordinates of a point in it are $(s',\bs{t})$ for
some $s'\in J^n$ and $\bs{t}\in\field[R]^{\dim M - n }$
	\end{Thm}

	For details concerning the construction of the local coordinates
$\{x^\mu\}$ in theorem~\ref{Thm5.1}, the reader is referred
to~\cite{bp-Frames-general,bp-NF-D+EP}

	From~\eref{5.2} an immediate observation
follows (see~\cite[sect.~5]{bp-Frames-general}): strong normal frames always
exist at every single point ($n=0$) or/and along every $C^1$ injective path
($n=1$). Besides, these are the \emph{only cases} when normal frames
\emph{always exist} because for them ~\eref{5.2} is identically valid.  On
submanifolds with dimension greater than or equal to two normal frames exist
only as an exception if (and only if)~\eref{5.2} holds. For $n=\dim M$
equations~\eref{5.2} express the flatness of the corresponding linear
connection.

	If on $U$ exists a frame $\{e_i\}$ normal for $\nabla$, then all
frames  $\{e'_i=A_i^je_j\}$ which are normal over $U$ can easily be
described: for the normal frames, the matrix $A=[A_i^j]$ must be such that
$E_\mu(A)|_U=0$ for some (every) frame $\{E_\mu\}$ over $U$ in $T(M)$.


\section {Electromagnetic potentials}
\label{Sect3}

	Recall~\cite{L&L-2,Drechsler&Mayer}, classical electromagnetic
field is described via a real 1\ndash form $A$ over a 4\ndash dimensional
real manifold $M$ (endowed with a Riemannian metric $g$ and) representing
the spacetime model and, usually, identified with the Minkowski space
$M^4$ of special relativity or the Riemannian space $V_4$ of general
relativity.%
\footnote{%
The particular choice of $M$ is insignificant for the following.%
}
The electromagnetic field itself is represented by the two\ndash form
$F=\od A$, where ``$\od$'' denotes the exterior derivative operator, with
local components (in some local coordinates $\{x^\mu\}$)
	\begin{equation}	\label{4-E.18}
F_{\mu\nu}
= - \frac{\pd A_\mu}{\pd x^\nu} +  \frac{\pd A_\nu}{\pd x^\mu} .
	\end{equation}

	As is well known, the electromagnetic field, the
Maxwell equations describing it,
 and its (minimal) interactions with other objects are
invariant under a gauge transformation
	\begin{equation}	\label{4-E.19}
A_\mu \mapsto A'_\mu
	= A_\mu + \frac{\pd \lambda}{\pd x^\mu}
	\end{equation}
or $A\mapsto A'=A+\od \lambda$, where $\lambda$ is a $C^2$ function. As is
almost evident, the electromagnetic field is invariant under simultaneous
changes of the local coordinate frame,
 $E_\mu=\frac{\pd}{\pd x^\mu}\mapsto E'_\mu=B_\mu^\nu E_\nu$
with $B_\mu^\nu:=\frac{\pd x^\nu}{\pd x^{\prime\,\mu}}$,
and a gauge transformation~\eref{4-E.19}:
	\begin{equation}	\label{4-E.20}
A_\mu \mapsto A'_\mu
	= B_\mu^\nu A_\nu + E'_\mu(\lambda)
	= B_\mu^\nu \Bigl(A_\nu + \frac{\pd \lambda}{\pd x^\nu} \Bigr) .
	\end{equation}
A simple calculation shows that under the transformation~\eref{4-E.20}, the
quantities~\eref{4-E.18} transform like components of an (antisymmetric)
tensor,
	\begin{equation}	\label{4-E.21}
F_{\mu\nu}\mapsto F'_{\mu\nu} = B_\mu^\sigma B_\nu^\tau F_{\sigma\tau}
	\end{equation}
due to which the 2-form $F$ remains unchanged, $F=\od A=\od A'$. Notice,
above $A'_\mu$ are \emph{not} the components of $A$ in $\{E'_\mu\}$ unless
$\lambda=\const$ while  $F_{\mu\nu}^{\prime}$ \emph{are} the components of
$F$ in $\big\{E^{\prime\mu}=\frac{\pd x^{\prime\mu}}{\pd x^\nu} \od
x^{\nu}\big\}$.

	Comparing~\eref{4-E.20} with~\eref{2.5}, we see that the former
equation is a special case of the latter one if we put in it
 $\dim\pi^{-1}(x)=1$, $x\in M$, $\Gamma_\mu=A_\mu$ and $A=\lambda$. That
simple observation reflects a known fundamental
result~\cite{Konopleva&Popov,Baez&Muniain,Drechsler&Mayer}:
	\emph{from geometrical view\ndash point the electromagnetic
potentials are coefficients of a covariant derivative (linear connection) (in
a given fields of bases -- \emph{vide infra}) in one\ndash dimensional real
vector bundle over the spacetime}.
	Precisely, let $(E,\pi,M)$ be one-dimensional real vector bundle over
the spacetime manifold $M$ and $\nabla$ be a linear connection in it.%
\footnote{~%
A one-dimensional vector bundle is called \emph{line bundle}.%
}
If $\{e\}$ is a 1\ndash vector frame in $E$ over $U\subseteq M$~%
\footnote{~%
We suppress the index~$1$ related to the frames in $E$, \ie we write $e$ for
$e_1$. However, if $u(x)\in\pi^{-1}(x)$, $x\in U$, we have to write \eg
 $u(x)=u^1(x)e(x)$ to distinguish $u(x)\in\pi^{-1}(x)$ from
$u^1(x)\in\field[R]$.%
}
and $\{E_\mu\}$ is a frame in the tangent bundle to $M$ over $U$, then the
coefficients $\Gamma_\mu$($\equiv\Sprindex[\Gamma]{1\mu}{1}$) of $\nabla$ in
$(\{e\},\{E_\mu\})$ are defined by (see~\eref{2.3})
	\begin{equation}	\label{30.1}
\nabla_{E_\mu} e = \Gamma_\mu e
	\end{equation}
and a non-degenerate change
	\begin{equation}	\label{30.2}
(\{e\},\{E_\mu\})\mapsto
(\{e'=\lambda e\},\{E'_\mu=B_\mu^\nu E_\nu\})
	\end{equation}
for $\lambda,B_\mu^\nu\colon U\to\field[R]$, with $\lambda$ being of
class $C^1$ and $\det[B_\mu^\nu]\not=0$, entails (see~\eref{2.4})
	\begin{equation}	\label{30.3}
\Gamma_\mu\mapsto
\Gamma'_\mu = B_\mu^\nu (\Gamma_\nu + E_\nu(\lambda)) .
	\end{equation}
Conversely, any geometrical object with components $\Gamma_\mu$ in
$(\{e\},\{E_\mu\})$ with transformation law~\eref{30.3} defines a unique
linear connection $\nabla$ with coefficients (coefficients' matrices)
$\Gamma_\mu$ via~\eref{30.1}. If we now specify $\{E_\mu\}$ as a coordinate
frame induced by local coordinates $\{x^\mu\}$, \ie
$E_\mu=\frac{\pd}{\pd x^\mu}$, we see that~\eref{30.3} and~\eref{4-E.20} are
identical up to the identification $\Gamma_\mu=A_\mu$, which completes the
proof of our assertion.

	A new moment in the geometrical treatment of the electromagnetic
potentials as coefficients of a linear connection is the clear meaning of the
gauge transformations~\eref{4-E.19} as transformation of the potentials under
the change
	\begin{equation}	\label{30.4}
(\{e\},\{E_\mu\})\mapsto (\{e'=\lambda e\},\{E_\})
	\end{equation}
corresponding only to a rescaling with factor $\lambda\colon
U\to\field[R]\setminus\{0\}$ of the single frame vector field $e$ of the
vector bundle frame $\{e\}$ in $\pi^{-1}(U)\subseteq E$ over a set
$U\subseteq M$.

	In this context, the different gauge conditions, which are frequently
used, find a natural interpretation as a partial fix of the class of frames
in the bundle space employed. For instance, any one of the gauges in the
table~\vpageref{4-TableOfGauges} corresponds to a class of frames for
which~\eref{4-E.20} holds for $B_\mu^\nu=\delta_\mu^\nu$, $\delta_\mu^\nu$
being the Kroneker deltas, and $\lambda$ subjected to a condition given in
the table.%
\footnote{%
Below $M$ is supposed  to be endowed with a Riemannian metric $g_{\mu\nu}$,
the coordinates to be numbered as $x^0,x^1,x^2$, and $x^3$, $x^0$ to be the
`time' coordinate, $\pd_\mu:=\pd/\pd x^\mu$, and $\pd^\mu:=g^{\mu\nu}\pd_\nu$
with $[g^{\mu\nu}]:=[g_{\mu\nu}]^{-1}$.%
}

	\begin{table}[ht!]	\label{4-TableOfGauges}
	\begin{minipage}{\textwidth}
	\begin{tabularx}{\textwidth}{lrll@{}}
Gauge  & Condition on $A$  & Condition on $\lambda$  & Condition on $\varphi$
\\ \hline
Lorentz
	& $\pd^\mu A_\mu =0$
	& $\pd^\mu\pd_\mu \lambda =0$
	& $\pd^\mu\pd_\mu \varphi = - \pd^\mu\pd_\mu \lambda $
\\
Coulomb\footnote{In this raw the summation over $k$ is from 1 to 3.}
	& $\pd^k A_k =0$
	& $\pd^k \pd_k \lambda =0$
	& $\pd^k \pd_k \varphi
		=- \pd^k \pd_k \lambda $
\\
Hamilton
	& $A_0=0$
	& $\lambda(x) = \lambda(x^1,x^2,x^3)$
	& $\varphi(x) = \varphi(x^1,x^2,x^3)$
\\
Axial
	& $A_3=0$
	& $\lambda(x) = \lambda(x^0,x^1,x^2,)$
	& $\varphi(x) = \varphi(x^0,x^1,x^2)$
\\ \hline
	\end{tabularx}
	\end{minipage}
	\end{table}
In the table~\vpageref{4-TableOfGauges} $\varphi$ is a $C^1$ function
describing the arbitrariness in the choice of $\lambda$, \ie if a gauge
condition is valid for $\lambda$, then it holds also for $\lambda+\varphi$
instead of $\lambda$.

	If an electromagnetic field exists on an open set $U\subseteq M$, then
its potentials admit an equivalent geometrical interpretation as 3\ndash
index coefficients of a linear transport along paths which is, in fact, the
parallel transport along paths for the linear connection whose coefficients
coincide with the field's potentials (see corollaries~\ref{Cor4.2}
and~\ref{Cor4.2new}).

	If one considers a free (pure) electromagnetic field, the bundle
space of the line bundle on which the field can be described as a linear
connection, remains undetermined.

	Suppose now an electromagnetic field exists on some submanifold $N$
of $M$ and $\dim N<\dim M$. (With some approximation such fields can be
realized.) In this case one cannot interpret the electromagnetic potentials
as coefficients of a linear connection on a line bundle $(E,\pi,M)$ over the
spacetime $M$. But such an interpretation is possible on the restricted
subbundle $(E,\pi,M)|_N=(\pi^{-1}(N),\pi|_{\pi^{-1}(N)},N)$ for which one can
repeat \emph{mutatis mutandis} the above considerations. However the
interpretation of field potentials as 3\ndash index coefficients of a linear
transport along paths can be retained. To demonstrate that, consider
one\ndash dimensional bundle $(E,\pi,M)$ and a linear transport $L$ along
paths in it such that the matrices of its coefficients satisfy the
condition
	\begin{equation}	\label{30.5}
\Mat{\Gamma}(s;\gamma)
= \Gamma_\mu(\gamma(s)) \dot\gamma^\mu(s)
\qquad
\text{for any $C^1$ path $\gamma\colon J\to N$ and $s\in J$}
	\end{equation}
for some matrix\ndash valued functions $\Gamma_\mu$ on $N$ in any frame
$\{e\}$ in $E$ over $N$ and frame $\{E_\mu\}$ in the tangent bundle space
over $N$. Otherwise the transport $L$ is completely arbitrary, \eg we can
require
	\begin{equation}	\label{30.6}
\Mat{\Gamma}(t;\beta)
= \Tilde{\Gamma}_\mu(\beta(t)) \dot\beta^\mu(t)
\qquad
\text{for any $C^1$ path $\beta\colon \Tilde{J}\to M$ and $t\in \Tilde{J}$}
	\end{equation}
for matrix\ndash valued functions $\Tilde{\Gamma}_\mu$ on $M$ such that
	\begin{equation}	\label{30.7}
\Tilde{\Gamma}_\mu(x) = \Gamma_\mu(x)
\qquad \text{for $x\in N$} .
	\end{equation}
If we identify $\Gamma_\mu(x)$, $x\in N$, with the electromagnetic
potentials $A_\mu$, then $A_\mu$ are 3\ndash index coefficients of any linear
transport $L$ along paths for which equation~\eref{30.5} holds. In invariant
terms, equation~\eref{30.5} can be rewritten as
	\begin{equation}	\label{30.8}
L^\gamma = \Psf^\gamma
\qquad \text{for $\gamma\colon J\to N$}
	\end{equation}
where $\Psf$ is the parallel transport along paths in
$(\pi^{-1}(N),\pi|_{\pi^{-1}(N)},N)$ whose coefficients are the
electromagnetic potentials (on $N$). Obviously, the condition~\eref{30.8}
does not define $L$ uniquely for paths which do not lie entirely in $N$.


\section {Equivalence principle in gravitation}
\label{Sect4}

	The primary role of the principle of equivalence is to ensure the
transition from general to special relativity. It has quite a number of
versions, known as weak and strong equivalence
principles~\cite[pp.~72--75]{Ivanenko&Sardanashvily-1985}, any one of which has
different, sometimes non\ndash equivalent, formulations. In the present paper
only the strong(est) equivalence principle is considered. Some of its
formulations can be found in~\cite{bp-PE-P?}.

	Freely speaking, an inertial frame for a physical system is a one in
which the system behaves in some aspects like a free one, \ie such a frame
`imitates' the absence (vanishment) of some forces acting on the system.
Generally inertial frames exist only locally, \eg along injective paths, and
their existence does not mean the vanishment of the field responsible for a
particular force. The best known examples of this kind of frames are for the
gravitational field. Below we rigorously generalize these ideas to classical
electrodynamics.

	In~\cite{bp-PE-P?} it was demonstrated that, when gravitational fields
are concerned, the inertial frames for them are the normal ones for the linear
connection describing the field and they coincide with the (inertial) frames
in which the special theory of relativity is valid. The last assertion is the
contents of the (strong) equivalence principle. In the present section, relying
on the ideas at the end of~\cite[sec.~5]{bp-PE-P?}, we intend to transfer
these conclusions to the area of classical electrodynamics.

	The normal frames are the mathematical concept
corresponding to/describing the physical one of inertial frames (of
reference). However, as we have seen in Sect.~\ref{Sect2}, frames normal for
a linear connection always exist at a given single point and/or along
(injective) path and on more general sets they exist only in some exceptional
case (see, e.g., theorem~\ref{Thm5.1}). This means that the (strong)
equivalence principle is valid at a given single point or path and on
submanifolds of the spacetime of dimension greater or equal to two it may be
true only for some special gravitational fields; in particular, on open sets
(which are submanifolds of dimension $\dim M=4$) it holds iff the linear
connection, describing the field, is curvature free, which physically is
interpreted as absence of the gravity field.

	The above conclusions have a general validity and concern non only
the general relativity but rather any gravitational theory in which the
gravitational field strength is identified with the coefficients of a linear
connection (in the tangent bundle over the spacetime).


\section {Equivalence principle in electrodynamics}
\label{Sect5}

	Consider a one-dimensional vector bundle $(E,\pi,M)$ over the
spacetime $M$ in which a classical electromagnetic field is described via a
linear connection $\nabla$ (or parallel transport $\Psf$) whose (3\ndash
index) coefficients coincide with the field's potentials $A_\mu$ in a pair
 $(\{e\},\{E_\mu\})$ of frames $\{e\}$ in $E$ and $\{E_\mu\}$ in the tangent
bundle space to $M$.

	A frame of reference will be called \emph{inertial} for an
electromagnetic field on a set $U\subseteq M$ if in it the field strength
vanishes on $U$, \ie if in it we have
	\begin{equation}	\label{50.1}
A'_\mu|_U = 0
	\end{equation}
as the field strength is (locally) identified with the electromagnetic
potentials. If an inertial frame exists on $U$, in it an electrically charge
particle (body) will behave in $U$ like a neutral one (if it is entirely
situated in $U$). Obviously, the mathematical object corresponding to an
inertial frame of reference on $U$ for an electromagnetic field is a frame (in
$E$) normal on $U$ for the linear connection describing it. The simple
observation of the
\emph{coincidence of inertial and normal frames is the contents of the
equivalence principle in classical electrodynamics}.%
\footnote{~%
Generally a frame of reference is a more complex concept than a coordinate
system or a field of bases in (some) bundle space of a vector bundle.
However, the other characteristics and properties of the physical concept of
a reference frame  are inessential in the context of the present
investigation.%
}
We can equivalently restate it as the assertion of coincidence of the
inertial frames and strong normal frames for the parallel transport along
paths describing the field (which is generated by or generates the linear
connection corresponding to the field).

	Comparing~\eref{4-E.18} with~\eref{5.3}, we get%
\footnote{%
Below we assume the Greek indices to run over the range 0, 1, 2, 3.%
}
	\begin{equation}	\label{4-E.23}
F_{\mu\nu} = R_{\mu\nu}(-A_0,-A_1,-A_2,-A_3) .
	\end{equation}
Thus, the electromagnetic field tensor  $F$ is completely responsible for the
existence of frames normal for the parallel transport $\Psf$
(theorem~\ref{Thm5.1}). For example, if $U$ is an open set, frames normal on
$U\subseteq M$ for $\Psf$ exist iff $F|_U=0$, \ie if electromagnetic field is
missing on $U$.%
\footnote{%
Elsewhere we shall prove that the components $F_{\mu\nu}$ completely describe
the curvature of $\Psf$ which agrees with the interpretation of $F_{\mu\nu}$ as
components of the curvature of a connection on a vector bundle in the gauge
theories~\cite{Konopleva&Popov,Drechsler&Mayer,Slavnov&Fadeev}. The general
situation is similar: the quantities~\eref{5.3} determine the curvature of
a transport with coefficients' matrix~\eref{4-5.1}.%
}
Also, if $N$ is a submanifold of $M$, frames normal on $U$ for $\Psf$ exist iff
in the special coordinates $\{x^\mu\}$, described in theorem~\ref{Thm5.1},
is valid $F_{\alpha\beta}|_U=0$ for  $\alpha,\beta=1,\dots,\dim N$.
In the context of~\cite[theorem~4.1]{bp-normalF-LTP}, we can say
that an electromagnetic field admits frames normal on $U\subseteq M$ iff
the corresponding to it linear transport  $\Psf$  is path\ndash independent on
$U$ (along paths lying entirely in $U$). Thus, if $\Psf$ is path\ndash dependent
on $U$, the field does not admit frames normal on $U$. This important result
is the classical analogue of the quantum effect, know as the Aharonov\ndash
Bohm effect~\cite{Aharonov&Bohm, Bernstein&Phillips}, whose essence is that
the electromagnetic potentials directly, not only through the field tensor
$F$, can give rise to observable physical results.

	Let us now turn our attention to the physical meaning of the normal
frames corresponding to a given electromagnetic field which is described, as
pointed above, via a parallel transport $\Psf$ along paths in 1\ndash
dimensional vector bundle over the space\ndash time $M$.

	Suppose $\Psf$ is Euclidean on a neighborhood $U\subseteq M$. As a
consequence of~\eref{4-E.23} and~\cite[theorem~5.1]{bp-normalF-LTP}, we have
$F|_U=\od A|_U=0$, \ie on $U$ the electromagnetic field strength vanishes and
hence the field is a pure gauge on $U$,
	\begin{equation}	\label{4-E.24}
A_\mu|_U = \frac{\pd f_0}{\pd x^\mu}\Big|_U
	\end{equation}
for some  $C^1$ function $f_0$ defined on an open neighborhood containing $U$
or equal to it. In a frame $\{e'\}$ normal on $U$ for $\Psf$ vanish the
2\ndash index coefficients of $\Psf$ along any path $\gamma$ in $U$:
	\begin{equation}	\label{4-E.25}
\Mat{\Gamma}'(s;\gamma) = A'_\mu(\gamma(s)) \dot\gamma^\mu(s) = 0
	\end{equation}
for every $\gamma\colon J\to U$ and $s\in J$. Using~\eref{4-E.24}, it is
trivial to see that any transformation~\eref{30.3} with
	\begin{equation}	\label{4-E.26}
\lambda = -f_0
	\end{equation}
transforms $A_\mu$ into $A'_\mu$ such that
	\begin{equation}	\label{4-E.27}
A'_\mu|_U = 0
	\end{equation}
(irrespectively of the frames $\{E_\mu\}$ and $\{E'_\mu\}$ in the tangent
bundle over $M$). Hence, by~\eref{4-E.25} the one\ndash vector frame
$\{e'=\e^{-f_0}e\}$ in the bundle space $E$ is normal for $\Psf$ on $U$.
Therefore in the frame $\{e'\},$ there  vanish not only the 2\ndash index
coefficients of $\Psf$ but also its \emph{3\ndash index} ones, \ie $\{e'\}$ is a
frame strong normal on $U$ for $\Psf$. Applying~\eref{4-E.20} one can verify,
\emph{%
all frames strong normal on a neighborhood $U$ for $\Psf$ are obtainable from
$\{e'\}$ by multiplying its vector $e'$ by a function $f$ such that
$\frac{\pd f}{\pd x^\mu}\bigr|_U=0$%
}, \ie they are $\{b\e^{-f_0}e\}$ with $b\in\mathbb{R}\backslash\{0\}$ as
$U$ is a neighborhood. Thus, every frame normal on a neighborhood $U$ for
$\Psf$ is strong normal on U for $\Psf$ and vice versa.

	So, in a frame inertial on $U\subseteq M$ for an electromagnetic
field it is not only a pure gauge, but in such a frame its potentials vanish
on $U$. Relying on the results obtained (see
also~\cite{bp-Frames-n+point,bp-Frames-path,bp-Frames-general}), we can
assert the existence of frames inertial at a single point and/or along paths
without self\ndash intersections for every electromagnetic field, while on
submanifolds of dimension not less than two such frames exist only as an
exception if (and only if) some additional conditions are satisfied, \ie for
some particular types of electromagnetic fields.


\section {Conclusion}
\label{Conclusion}

	In this paper we introduced normal an inertial frames for classical
electromagnetic field. The coincidence of these two types of frames
expresses the equivalence principle for that field. Generally this principle
is a provable theorem and it is always valid at any single point of along
given path (without selfintersections) as these are the only case when
normal frames for a linear connection always exist.

	The considerations of the equivalence principle in electrodynamics
were base on the interpretation of a classical electromagnetic field as a
linear connection or (the generating it or generated by it) parallel transport
along paths in one\ndash dimensional vector bundle over the spacetime.
Within the electrodynamics, \ie for a free electromagnetic field, that
bundle remains unspecified. However, if an interaction of electromagnetic
field and some other field is investigated, the bundle mentioned can be
identified or uniquely connected with a bundle (over the spacetime) whose
sections represent the latter field. Moreover, in such a situation the
equivalence principle can be used to justify the so\ndash called minimal
coupling (principle), \ie the description of the interaction of some field
with an electromagnetic one via a replacement of the ordinary partial
derivatives in the free Lagrangian of the former field with covariant ones
relative to the linear connection representing the latter field.

	At last, we would like to mention that the existence of a
normal/inertial frames on some subset does not generally imply vanishment of
the field on this set if it is not an open set.




\addcontentsline{toc}{section}{References}
\bibliography{bozhopub,bozhoref}
\bibliographystyle{unsrt}
\addcontentsline{toc}{subsubsection}{This article ends at page}

\end{document}